\documentstyle[12pt,epsf]{article}
\newlength{\dinwidth}
\newlength{\dinmargin}
\begin{document}
\def\thefootnote{\fnsymbol{footnote}}

\baselineskip18pt

\thispagestyle{empty}

\begin{flushright}
\begin{tabular}{l}
CERN-TH.7085/93\\
FTUOV-100/93\\
\end{tabular}
\end{flushright}

\vspace*{1.4cm}

{\vbox{\centerline{{\Large{\bf A COSMOLOGICAL INTERPRETATION OF DUALITY
}}}}}

\vskip25pt

\centerline{M.\,A.\,R. Osorio\footnote{E-mail address:
    {\tt OSORIO@PINON.CCU.UNIOVI.ES}}}

\vskip8pt
\centerline{{\it Departamento de F\'{\i}sica}}
\centerline{{\it Universidad de Oviedo}}
\centerline{{\it Avda. Calvo Sotelo 18}}
\centerline{{\it E-33007, Oviedo, Spain}}

\vskip8pt

\centerline{{\it and}}

\vskip8pt

\centerline{M.\,A. V\'azquez-Mozo\footnote{E-mail address: {\tt
      VAZQUEZ@CERNVM.CERN.CH}} \footnote{On leave of absence from
Dept. F\'{\i}sica Te\'orica C-XI, Universidad Aut\'onoma de Madrid,
E-28049 Madrid, Spain.}}

\vskip8pt
\centerline{{\it Theory Division, CERN}}\vskip2pt
\centerline{{\it CH-1211 Geneva 23}}\vskip2pt
\centerline{{\it Switzerland}}

\vskip .25in

\baselineskip18pt
\indent

We study the cosmological meaning of duality symmetry by considering a
two dimensional model of string cosmology. We find that as seen by an
internal observer in this universe, the scale factor rebounds at the
self-dual length. This rebound is a consequence of the adiabatic
expansion. Furthermore, in this situation there are four
mathematically different
scenarios which describe physically equivalent universes which are in fact
undistinguishable. We also stress that $R$-duality suffices to prove that
all the possible evolutions present a maximun temperature.

\baselineskip18pt

\vspace*{24pt}

\noindent
CERN-TH.7085/93\\
FTUOV-100/93\\
November, 1993

\setcounter{page}{0}

\newpage

Duality symmetry is the most important stringy symmetry from many
points
of view (for a review see \cite{Schwarz}). Let us suppose we have a string
propagating in a target space $\mbox{\bf R}^{d-1}\times S^{1}$
where we have set the radius of the compactified dimension equal to
$R$.
It is a well known fact that every correlation function $A(1,\ldots,N)$
can be written
as a topological expansion in the string coupling constant
\begin{equation}
A(1,\ldots,N)=\sum_{g=0}^{\infty} g_{st}^{\,2(g-1)}
A_{g}(1,\ldots,N)\:,
\end{equation}
where $A_{g}$ is the correlator at fixed genus. Duality symmetry
means that $A(1,\ldots,N)$ as a function of $R$ and $g_{st}$ is
invariant
under the replacement \cite{alvarez-osorio-d40}
\begin{equation}
R\longrightarrow \frac{\alpha^{'}}{R}\;, \hspace{1cm}
g_{st} \longrightarrow \frac{\sqrt{\alpha^{'}}}{R}g_{st}
\label{duality}
\end{equation}
together with an interchange between the momentum and the winding modes
of the external states.
In other words, we are unable to distinguish between small and large $R$
provided we change the string coupling constant properly. Since this
symmetry is preserved by the whole topological expansion, we have
that if broken it cannot be within the realm of string perturbation
theory.

Since any string scattering experiment is unable to tell us whether we
are living
in a universe with size $R$ and string coupling constant $g_{st}$ or in a
universe with the dual values, it has been argued by some authors
\cite{Gross-Veneziano} that this defines in fact a minimum measurable
length
at the self-dual distance $\sqrt{\alpha^{'}}$. This has led to a
modification of the Heisenberg uncertainty principle in order to
include
this new feature implied by String Theory
\begin{equation}
\Delta x \sim \frac{\hbar}{\Delta E} + \alpha^{'} \Delta E\;.
\end{equation}
Thus, independently of the value of $\Delta E$, the uncertainty in the
position $\Delta x$ is above some minimum value of the order of
$\sqrt{\alpha^{'}}$.

The implications of duality symmetry are very important from a
phenomenological
point of view (see for example \cite{phenomenology}). However, in
this work we are only
concerned with the meaning of duality symmetry in the
cosmological
context. This subject has been already investigated in a number of
works
\cite{brandenberger-vafa,Tseytlin2,tseytlin-vafa,tseytlin1,var-1,var-2}.
In ref.
\cite{brandenberger-vafa} it was argued that duality symmetry together
with
the existence of a Hagedorn temperature for the string gas filling
the universe would imply that the size of the universe as a function of
the cosmic time had to rebound at the selfdual size. In this picture
one starts with a universe with all spatial dimensions compactified and
of
the order of the Planck length and ends up with a universe in which
only
three of the spatial dimensions have grown above the self-dual size
while the others remain at the Planck length
scale. In ref. \cite{var-2} the numerical solutions for
a two-dimensional space-time filled with a gas of two-dimensional
critical
strings was studied and it was found that there exists a class
of solutions for
which the scale factor decreases from infinity to zero value (see Fig. 1).
The existence
of this type of solutions seems to be contradictory with the scenario
depicted by Brandenberger and Vafa in \cite{brandenberger-vafa}, since
no dynamical rebound of the scale factor is seen.

In this brief letter we give an interpretation for the
results
found in \cite{var-2} that are compatible with the image of the
scale factor rebounding at the self-dual length. Although our
discussion
will be concentrated on a two-dimensional model, the conclusions extracted
can be extrapolated to other scenarios with or without Hagedorn
temperature.

Let us suppose we have a string moving in a target space
$\mbox{\bf R}^{d-1}\times S^{1}$ with metric
\begin{equation}
ds^{2}=ds_{d-1}^{2}+R^{2}d\theta^{2}\;,
\label{metric}
\end{equation}
where $ds^{2}_{d-1}$ is the line element in $\mbox{\bf R}^{d-1}$ and
$\theta\in[0,2\pi)$. Let also the
parameter $R$ become dynamical, i.e., a function $R(t)$ of the cosmic
time. It is easy to see, and certainly surprising, that the Brans-Dicke
action
\begin{equation}
S=\int d^{d}x \sqrt{-g}\left[\Phi(R-2\Lambda)-\frac{\omega}{\Phi}
\nabla_{\mu}\Phi\nabla^{\mu}\Phi\right]\:,
\label{Brans-Dicke}
\end{equation}
endowed with the {\it ansatz} (\ref{metric}) and a space-independent
Brans-Dicke field $\Phi(t)$  is invariant
under the following substitution \cite{Tseytlin2,Veneziano2}
\begin{equation}
R(t)\longrightarrow \frac{\alpha^{'}}{R(t)}\:, \hspace{1cm}
\Phi(t) \longrightarrow \frac{R(t)^{2}}{\alpha^{'}} \Phi(t)\:,
\label{BD-symmetry}
\end{equation}
However it is worth stressing that this symmetry, although closely
resembles duality in string theory upon the
identification of the square root of the string coupling constant with
the inverse of the dilaton field, it is not the same thing
than (\ref{duality}). The reason is
that duality is a symmetry between two backgrounds. In the case
discussed before these vacua are two copies of
$\mbox{\bf R}^{d-1}\times S^{1}$ with
dual values for the radius of the compactified dimension. In our case
$R(t)$ is no longer a constant, and the background space is
the product of $\mbox{\bf R}^{d-2}$ by a kind of {\it trumpet}
with topology $\mbox{\bf R}\times S^{1}$. Duality symmetry will relate this
background manifold with other background manifold which, in general,
will
not be that obtained by substituting $R(t)\rightarrow \alpha^{'}/R(t)$.
There is however a way in which (\ref{duality}) and (\ref{BD-symmetry})
can
be related. If the variation of $R(t)$ is not very wild,
physically, the string at any instant of time would {\it think} that
the
background space in which it is moving has a constant radius for the
compatified dimension. This implies that, at any instant of time $t$,
all
physical observables will enjoy duality symmetry (\ref{duality}) with
$R=R(t)$ which is exactly the same symmetry that the Brans-Dicke action
has. This means, in particular, that the matter action to be coupled to
(\ref{Brans-Dicke}) will also be invariant under (\ref{BD-symmetry})
and
then all physical quantities derived from this action functional will
present the same symmetry. It is worth remarking that in this sense
duality symmetry has been present in Physics since the sixties
waiting for a theory (String Theory) that could provide a matter
action with the same symmetry. This matter action in the case at hand
can
be constructed from the one-loop Helmholtz free energy in
the way described in \cite{var-2} (cf. also \cite{tseytlin-vafa}).

In \cite{var-2} we studied numerically the cosmological
solutions
for a two-dimensional universe $S^{1}\times\mbox{\bf R}$ filled with string
matter.
As we said, there are solutions that
describe a universe contracting from infinite size down to $R=0$. From
a mathematical point of view we find that duality symmetry does not
imply the existence of a dynamical rebound at the self-dual radius.
The  only meaning of duality from a mathematical point of view is that if we
find a solution described by $R(t)$ and $\Phi(t)$ the dualized
functions
are also solutions of the system of differential equations. However these
solutions present a strange feature, at least strange from a quantum
field theoretical point of view: although the contraction is adiabatic,
as the size of the universe monotonously goes to zero the temperature
reaches a maximum and then decreases and is actually zero for zero size.
Comparing $T(t)$ with $R(t)$ one finds that the maximum is gotten when
the size is the self-dual point of the duality transformation. What we really
then have is that the number of excited degrees of freedom below the
self-dual point does not increase; on the contrary it decreases and
we have the same number at $R=0$ than at $R=\infty$. These are the
workings of the equation of state which differ from the field theoretical
case $\rho = p$. The entropy of a system measures, roughly speaking, the
number of degrees of freedom that can be excited at a given temperature.
For an entropy function which enjoys duality as a function of the volume,
it implies the absence of excitable degrees of freedom at small radius.
So duality
supports the suspicion that some authors have had about the scarce number of
degrees of freedom that can be excited below the Planck length
(see for example \cite{scarce}).

Let us elaborate further the analysis of duality symmetry from a purely
physical point of view. We  have
to consider an observer inside our two-dimensional universe performing
a set of experiments in order to measure the size $R$, the temperature
$\beta^{-1}$ and the strength of the coupling $g_{st}$ (or equivalently
$\Phi$). Of course we have to assume that the time required by these
experiments is much less than the characteristic time of expansion, in
order to regard the universe as locally static. In that case, what
duality symmetry would imply is that our one-dimensional
experimentalist
is unable to decide whether he/she is living in a universe with
radius $R$ and a gravitational constant given by $g_{st}$ or in the
dual universe (let us remind that the inverse temperature $\beta$ is
not affected by duality symmetry). Let us assume that he/she chooses one
of the two possible branches arbitrarily; say branch 1 in Fig. 1 defined
by $t \leq t_{sd}, \, R(t) \geq \sqrt{\alpha '}$.
Since the universe that we are considering (branches 1 and 4 of Fig. 1)
has a monotonously decreasing
scale factor, for every time $t_{0}$ there is a time $t_{0}^{*}$ such
that
\begin{equation}
R(t_{0}^{*})= \frac{\alpha^{'}}{R(t_{0})}\;.
\end{equation}
The question now is whether the internal observer is able to decide
if the state of the universe at $t_{0}$ is different from that in
$t_{0}^{*}$. The only way in which this can be accomplished is if
both the field $\Phi$ and the inverse temperature $\beta$ at time
$t_{0}^{*}$ are different from the dual values of $\Phi(t_{0})$ and
$\beta(t_{0})$. Let us consider the case of the inverse temperature
$\beta$. To show that $\beta(t_{0})=\beta(t_{0}^{*})$ it is enough
the fact that the entropy $S(\beta,R)$ is a dual function, i.e.,
\begin{equation}
S(\beta,R)=S\left(\beta,\frac{\alpha^{'}}{R}\right)\;,
\end{equation}
and that the scale factor $R(t)$ is a monotonous function of time.
In fact, since $R(t)$ is a monotonous function, we can parametrize
the evolution of our universe by $R$ instead of $t$. This means that
the entropy is a function of $R$, but since the expansion is adiabatic
(i.e., $S(R)$ is constant), we have
\begin{equation}
S(R)=S[\beta(R),R]=S\left[\beta\left(\frac{\alpha^{'}}{R}\right),
\frac{\alpha^{'}}{R}\right]=S\left(\frac{\alpha^{'}}{R}\right)\;.
\end{equation}
But, because of $R$-duality we have also
\begin{equation}
S\left(\frac{\alpha^{'}}{R}\right)= S\left[\beta\left(
\frac{\alpha^{'}}{R}\right), R\right]\;.
\end{equation}
This, together with the single-valuedness and the monotonous character
of the entropy with respect
to $\beta$ (the second is the result of the fact that the specific
heat at constant volume is positive) implies that
\begin{equation}
\beta(R)=\beta\left(\frac{\alpha^{'}}{R}\right)\;,
\end{equation}
so the temperature at $t_{0}$ and $t_{0}^{*}$ are exactly the same.
Duality also holds for the internal energy and then the whole thermodynamics
for the universe of size $R$ is indistinguishable from that at $\alpha^{'}/R$.
By the way, it is easy to see that this relation implies that the temperature
has a extreme at $R=\sqrt{\alpha^{'}}$ (indeed a maximum).
Then the only way in which the internal observer could distinguish
the universe in $t_{0}$ from the universe in $t_{0}^{*}$ is if
\begin{equation}
\Phi(t_{0}^{*})\neq \frac{R^{2}(t_{0})}{\alpha^{'}} \Phi(t_{0})\;.
\label{ineq}
\end{equation}
Then we have to address the problem of measuring $\Phi$ (or
equivalently
$g_{st}$). Since we are working at one (thermal) loop level, we are
neglecting any interaction between strings. This means that in this
approximation the strings are free and then there is no hope of
measuring
the string coupling constant, so $g_{st}$ is not an observable for our
internal observer. The question about whether (\ref{ineq}) holds
seems to be irrelevant at one loop. It would be necessary to make the
computation
to two loops in order $g_{st}$ to be a measurable parameter.
Nevertheless it is well known that static duality including the transformation
of the dilaton field arises in the sigma model
even computing at world-sheet tree level \cite{buscher}. To
better understand  the  situation, let us summarize  the way in which
the observer would describe the evolution
of his/her universe. We have assumed that in the initial measurement at
$t_{0}$ the observer has chosen one of the two possible branches (say,
$R\geq \sqrt{\alpha^{'}}$). Then, at successive instants
$t_{1},\ldots,t_{n-1}$ he/she performs different measurements of $R$
and plot them.
When the scale factor as measured from outside decreases below  the
self-dual distance (i.e., when $t>t_{sd}$) the result of the measurements
will be
exactly the same as those for a given $t^{'}<t_{sd}$ and this is
interpreted by our experimentalist as the indication that the scale  factor
has suffered a rebound after reaching the self-dual size.
The observer, travelling forward in time, has actually jumped from the
solution coming from infinity to the dual one, i.e. $\alpha^{'}/R(t)$,
heading back to infinity (he has jumped from branch 1 to branch 2 in Fig 1).
If, after jumping, he/she tries to put to test Einstein-Brans-Dicke equations
the result will be that they hold, provided that the dilaton is changed
from $\Phi$ to
$R^2/\alpha^{'}\Phi$. Indeed these equations are the conditions up to the
lowest order in $\alpha'$ for the sigma model to define a conformal theory.
Both branches (1 and 2 in Fig. 1) differ because of the initial conditions
but these conditions are dual between them. The branches are glued together
at the
self-dual point in a non smoothly way; the first derivative of the
scale factor at $t_{sd}$ jumps by a finite quantity.

We have seen that the observer travelling forward in time has the possibility
of interpreting his/her Universe either as coming from infinity and going to
a cold end without size or coming from infinity and bouncing back to expand
at the Planck length. Once he/she is in a universe which expands
it is possible by inverting
time to go back to the universe of self-dual size (of course
theoretically). After moving a lapse back in time from $t=t_{sd}$
the observer also have
two different branches to choose. He/she can either go back
to infinite size (branch 1 in Fig.1) or take
another path smoothly going to zero size (branch 3 in Fig.1).
It is clear that a solution
$R(t)$ and its dual $\alpha^{'}/R(t)$ define four physically equivalent
scenarios given by taking together branch 1 and 4 or 3 and 2
or 1 and 2 or 3 and 4 (see Fig. 1).
This is precisely the number of different ways we can represent
the solitonic contribution to the partition function computed in a
static background \cite{note}: as a sum over a pair of windings,
as a sum over windings and momenta (we have two
possibilities of this kind), and as a sum over a pair of momenta, all of
them are related by applying Poisson summation formula to
the solitonic sum in the partition function.
Each time our observer jumps at the self-dual distance, he/she
performs a Poisson summation to redefine the eigenstates of the position
operator.

An important consequence of the combined effect of duality and the
adiabatic expansion is the fact that for every solution of those
described
in \cite{var-2} we find a maximum value of the temperature in the
universe.
In particular, for those solutions which go through the self-dual size
this
maximum can be easily seen to be located at $t_{sd}$ \cite{var-2}. The
absence of a Hagedorn temperature in this case makes the analysis more
transparent than for critical strings. However we see the similitude
with the case discussed in \cite{brandenberger-vafa}. There, since all
the spatial dimensions were compactified, the Hagedorn temperature was the
maximum temperature of the universe and was reached at the self-dual
size.
The rebound of the scale factor was there, as  here, a consequence of
the change
between momentum and winding modes to describe localized states. In our
scenario there is a maximum temperature although we do not have a
Hagedorn one.
This indicates that the scenario depicted
in \cite{brandenberger-vafa} is independent of the existence of a
Hagedorn temperature, and can be extended to other situations as
that studied in \cite{var-2}.

\section*{Acknowledgements}

We thank E. Alvarez and T. Ort\'{\i}n for reading the
manuscript. We also thank T. Ort\'{\i}n for some interesting comments.
We wish to thank the people and village of
Salas (Asturias, Spain) for providing us with a small but beautiful
and green piece of Universe where the main ideas of this paper were
conceived. M.A.V.-M. thanks CERN Theory Division for hospitality.
The work of M.A.V.-M. has been partially supported by a Comunidad de
Madrid Fellowship.

\newpage

\begin{figure}[t]
\epsffile{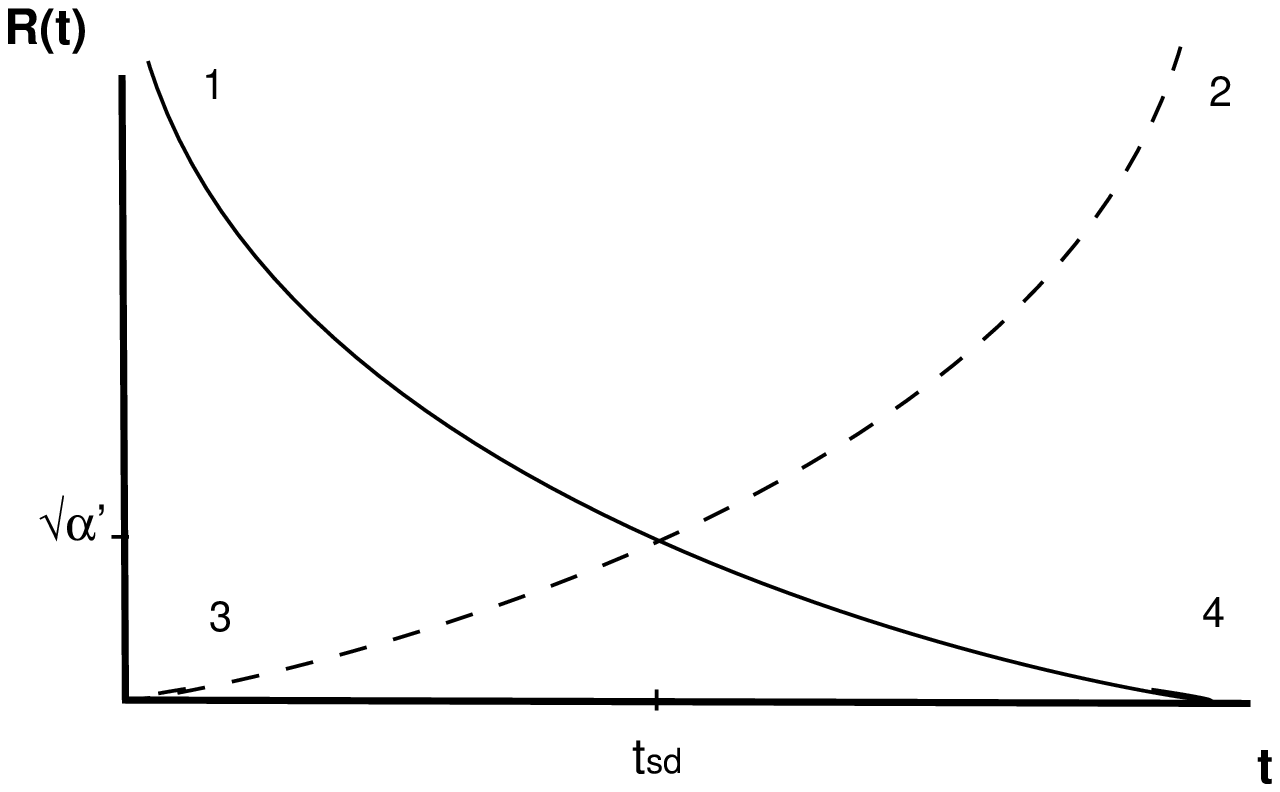}
\vspace*{-14cm}
\caption{Scale factor $R(t)$ versus time showing the four possible
branches.}
\end{figure}


\begin{thebibliography}{99}
\bibitem{Schwarz} J. Schwarz, {\it Spacetime Duality in String Theory},
in
``Elementary Particles and the Universe. Essays in honor of Murray
Gell-Mann'', J. Schwarz ed. Cambridge 1991.

\bibitem{alvarez-osorio-d40} E. Alvarez and M. Osorio, Phys. Rev. {\bf
D40} (1989) 1150.

\bibitem{Gross-Veneziano} D. Gross, {\it Strings and Unification}, talk
given at the 24th Int. Conf. on High Energy Physic, Munich, 1988.
\\
G. Veneziano, {\it Physics with a fundamental length}, in: ``Physics and
Mathematics of Strings: Memorial volume for V. Knizhnik'',
L. Brink, D. Friedan and A. Polyakov eds. World Scientific, Singapore 1990.

\bibitem{phenomenology} A. Font, L. Iba\~nez, D. L\"ust and
F. Quevedo, Phys. Lett. {\bf 245B} (1990) 401.
\\
M. Cvetic, A. Font, L. Iba\~nez, D. L\"ust and F. Quevedo, Nucl. Phys.
{\bf B361} (1991) 194.

\bibitem{brandenberger-vafa} R. Brandenberger and C. Vafa, Nucl. Phys.
{\bf B316} (1988) 391.

\bibitem{Tseytlin2} A. Tseytlin, Mod. Phys. Lett. {A} (1991) 1721.

\bibitem{tseytlin-vafa} A. Tseytlin and C. Vafa, Nucl. Phys. {\bf B372}
(1992) 443.

\bibitem{tseytlin1} A. Tseytlin, Class. Quant. Grav. {\bf 9} (1992) 979.

\bibitem{var-1} M. Osorio and M. V\'azquez-Mozo, {\it Variations on
Kaluza-Klein Cos\-mo\-lo\-gy}, Mod. Phys. Lett. A to appear.

\bibitem{var-2} M. Osorio and M. V\'azquez-Mozo, {\it String Variations
on Kaluza-Klein Cosmology}, Mod. Phys. Lett. A to appear.

\bibitem{Veneziano2} G. Veneziano, Phys. Lett. {\bf 265B} (1991) 287.

\bibitem{scarce} J. Atick and E. Witten, Nucl. Phys. {\bf B310} (1988) 291.
\\
M. Osorio and M. V\'azquez-Mozo, Phys. Lett. {\bf 280B} (1992) 21.
\\
M. Osorio and M. V\'azquez-Mozo, Phys. Rev. {\bf D47} (1993) 3411.
\\
L. Susskind, {\it Some speculations about black hole entropy in String
Theory}, Rutgers preprint RU-93-44, 1993 (hep-th/9309145).

\bibitem{buscher} T. Buscher, Phys. Lett. {\bf 201B} (1988) 466.

\bibitem{note} M. Osorio, Mod. Phys. Lett. {\bf A5} (1990) 1779.

\end{thebibliography}
\end{document}